\documentstyle[sprocl]{article}

\def\be{\begin{equation}}
\def\ee{\end{equation}}
\def\bea{\begin{eqnarray}}
\def\eea{\end{eqnarray}}
\def\ltap{\ \raisebox{-.4ex}{\rlap{$\sim$}} \raisebox{.4ex}{$<$}\ }
\def\gtap{\ \raisebox{-.4ex}{\rlap{$\sim$}} \raisebox{.4ex}{$>$}\ }

\begin{document}
{\hfill Ref. SISSA 66/99/EP}

{\hfill June 6, 1999}
\title{Neutrino Mixing and Oscillations in 1999 and Beyond
\footnote{ Invited talk given at the International Workshop on Weak 
           Interaction and Neutrinos, 
           January 25 - 30, 1999, Cape Town, South Africa
           (to be published in the 
           Proceedings of the Workshop).}  
}
\vskip 0.8cm
\author{S. T. Petcov \footnote{Also at: Instutute of Nuclear Research 
and Nuclear Energy, Bulgarian Academy of Sciences, 1789 Sofia, Bulgaria.}}

\address{Scuola Internazionale Superiore di Studi Avanzati, I-34014 Trieste,
Italy, and\\
Istituto Nazionale di Fizica Nucleare, Sezione di Trieste, I-34014
Trieste, Italy}


\maketitle\abstracts{The interpretation of 
the existing experimental eveidences for 
oscillations of neutrinos in schemes with 
three and four neutrino mixing is reviewed.
Forms of the
lepton mixing matrix allowed by 
the neutrino oscillation data
are considered. The possible neutrino 
mass spectra compatible with the 
observations 
are analyzed.
The possibility
to obtain information about the neutrino
mass spectrum from the future
$^{3}$H $\beta-$decay
and $(\beta \beta)_{0\nu}-$decay experiments
is also considered.}
\section{Introduction}
   At present there exist strong 
evidences that the flavour neutrinos,
$\nu_e$ and $\nu_{\mu}$ take part in 
neutrino oscillations \cite{Pont57} 
(see also \cite{MNS62,Pont67}) or undergo transitions in matter 
\cite{MSW} into neutrinos of different type.
They come primarily from the experiments 
with solar \cite{Bellotti,Suzuki} 
and atmospheric neutrinos \cite{Suzuki,Litchf}.
Indications for neutrino
oscillations have been obtained also in the 
LSND neutrino oscillation experiment \cite{LSND}:
the anomalous events observed in this experiment
 can be interpreted as being due
to small mixing 
$\bar{\nu}_{\mu} \leftrightarrow \bar{\nu}_{e}$
oscillations.

  These evidences suggest that
neutrinos have nonzero
masses and that lepton mixing is present 
in the weak charged lepton current.
If definitely established, 
the existence of nonzero 
neutrino masses and of lepton mixing
will have profound implications for our understanding
of the elementary particle interactions. 
It entails a number of important and not easy to answer questions
(see, e.g., \cite{BPet87}) as well. 
Here we give a possible short list. 
i) The number of massive neutrinos can be equal to, or be greater
than, the number of flavour neutrinos. 
Which of the two possibilities is realized, 
in other words,  are there sterile neutrinos, $\nu_s$?
What is the number
of neutrinos with definite mass? What is it determined by?
ii) The neutrino oscillation data does not permit to 
determine the absolute values of the neutrino masses.
What are they, i.e., what is the neutrino mass spectrum? 
iii) What are the values of the elements of 
of the lepton mixing matrix? Does the lepton mixing matrix
contain nontrivial
CP-violation phases? Are there  CP- violation
effects in neutrino oscillations?
iv) Neutrinos with definite mass 
can be Dirac or Majorana particles. 
Which of the two possibilities is realized and why?
v) Massive neutrinos
and lepton mixing imply that
the additive lepton charges - the electron $L_e$, the muon $L_{\mu}$ and
the tauon $L_{\tau}$, are not conserved by the
elementary particle interactions.
 Do lepton number non-conserving processes other 
than neutrino oscillations, as like $\mu^{-} \rightarrow e^{-} + \gamma$,  
$\mu^{-} \rightarrow e^{-} + e^{+} + e^{-} $, etc. 
exist at observable level?        
This list can be continued. It is clear 
that by proving the existence
of nonzero neutrino mass and        
of lepton mixing one would establish the
existence of a whole ``new world'' in elementary
particle physics. There are strong evidences that
this ``new world'' indeed exists and 
we are just beginning to explore it. 

    In the present article we shall review  
the evidences for oscillations of neutrinos. 
The interpretation of the  
solar and atmospheric neutrino data 
in a scheme with three-neutrino mixing
will be considered.
We shall discuss briefly the 
new effect of maximal enhancement of the 
transitions in the Earth 
of the solar and atmospheric neutrinos
which cross the Earth core on the way to the detectors.
Further, results obtained 
assuming four neutrino mixing and including 
the LSND indications for oscillations
in the analysis of the data will be reviewed. 
Rather simple patterns of lepton mixing 
emerge from these analyses. 
We will consider next    
the possibility 
to obtain information about the neutrino
mass spectrum from the future 
$^{3}$H $\beta-$decay 
and $(\beta \beta)_{0\nu}-$decay experiments.

\section{The Data or the ``Initial'' and the`` Boundary'' 
Conditions of the Analysis}

\indent  We have strong indications for vacuum 
$\nu_e \leftrightarrow \nu_{\mu, \tau}$ 
oscillations (VO),
or MSW $\nu_e \rightarrow \nu_{\mu, \tau}$ or 
$\nu_e \rightarrow \nu_{s}$  
transitions, of the solar neutrinos
from the mean event rate
solar neutrino data (see, e.g., \cite{Hata,BKS98}). 
If one uses the data
as published by the Homestake, SAGE, GALLEX and 
Super-Kamiokande collaborations 
in the analyses of the VO or MSW hypotheses, one finds that 
i) the $\nu_e \leftrightarrow \nu_{s}$ oscillations in vacuum,
ii) the $\nu_e \rightarrow \nu_{s}$ large 
mixing angle (LMA) MSW transitions
of solar neutrinos,
as well as iii) a universal energy-independent    
suppression of the solar neutrino flux,
are disfavored (in some of the cases - strongly) 
by the data \cite{KP94,KP96,KP4,KP97,Hata,BKS98}. 
This conclusion is based, to large extent,
on the result of the Homestake experiment when compared
with the results of the other solar neutrino experiments. 
If, e.g., the systematic error in the $^{37}$Ar production rate,
reported by the Homestake collaboration, is arbitrarily 
increased by a factor of $\sim 3$, the above results change and,
for instance, the hypothesis of a constant suppression 
of the solar neutrino flux by a factor of $\sim$0.5
becomes no longer strongly disfavored by the data.
However, we do not see at present any reasons for 
changing the value of the systematic uncertainty
in the Homestake data, given by the Homestake collaboration,
and will not consider further
the indicated three disfavored possibilities. 
We will not discuss also the possible solution
of the solar neutrino problem based 
on the hypothesis of existence of new neutrino flavour changing
and neutrino flavour conserving but flavour non-symmetric 
neutral current interactions \cite{FCNC}, or  
of violation of the weak equivalence principle \cite{GravOsc}, etc.
Both have difficulties
in explaining the atmospheric 
neutrino data \cite{LipLusign}.
We will be interested in the simplest solutions 
of the solar neutrino problem, which
can be incorporated naturally in schemes
providing explanation 
of the atmospheric neutrino anomalies 
as well. Following these guiding rules 
we are left with just four generic possibilities: 
the large mixing vacuum oscillations 
$\nu_e \leftrightarrow \nu_{\mu, \tau}$,
the small and the large mixing MSW  
$\nu_e \rightarrow \nu_{\mu, \tau}$ transitions and
the small mixing (SMA) MSW $\nu_e \rightarrow \nu_{s}$ 
transitions.

     According to the recent analyses \cite{Hata,BKS98}, 
the two-neutrino  
$\nu_e \leftrightarrow \nu_{\mu, \tau}$ 
oscillations in vacuum
provide a description (at 95\% C.L.) 
of the solar neutrino data for values of the two vacuum
oscillation parameters, $\Delta m^2$ and $\sin^22\theta$,
belonging approximately to the region:
$$5.0\times10^{-11} {\rm eV}^2\ltap \Delta m^2 \ltap
5.0\times 10^{-10} {\rm eV}^2,~~\eqno(1a)
$$
$$
0.65 \ltap \sin^22\theta \leq 1.0.~~~\eqno(1b)
$$
\noindent The SMA and LMA MSW $\nu_e \rightarrow \nu_{\mu, \tau}$ 
transition solutions 
take place for values of  
$\Delta m^2$ and $\sin^22\theta$
from the intervals 
$$4.0\times10^{-6}~ {\rm eV}^2\ltap \Delta m^2 \ltap
9.0\times 10^{-6}~ {\rm eV}^2,~~\eqno(2a)
$$
$$
10^{-3} \ltap \sin^22\theta \leq 10^{-2}.~~~\eqno(2b)
$$
\noindent and
$$2.0\times10^{-5}~ {\rm eV}^2\ltap \Delta m^2 \ltap
2.0\times 10^{-4}~ {\rm eV}^2,~~\eqno(3a)
$$
$$
0.65 \ltap \sin^22\theta \leq 1.0.~~~\eqno(3b)
$$
\noindent while the SMA MSW
$\nu_e \rightarrow \nu_{s}$ transition solution
is realized for    
$$3.0\times10^{-6}~ {\rm eV}^2\ltap \Delta m^2 \ltap
8.0\times 10^{-6}~ {\rm eV}^2,~~\eqno(4a)
$$
$$
1.5\times 10^{-3} \ltap \sin^22\theta \ltap 1.2\times 10^{-2}.
~~~\eqno(4b)
$$    
     Although these results are obtained utilizing the standard
solar model predictions of
ref. \cite{BP98} for the fluxes of
the $pp$, $pep$, $^{7}$Be, $^{8}$B and CNO neutrinos, 
they are rather stable
with respect to variation of 
the fluxes within their estimated
uncertainty ranges (see, e.g., \cite{BKS98,KP4,KLP96}) and so is
the existence of the four generic solutions.
Moreover,
the relative magnitudes of 
$\sin^22\theta$ and of 
$\Delta m^2$ for the four solutions 
remain unchanged with respect to such variations: 
we always have, e.g., 
$\Delta m^2_{VO} \ll \Delta m^2_{SMA} < \Delta m^2_{LMA}$  
and $\sin^22\theta_{SMA} \ll \sin^22\theta_{LMA} \sim \sin^22\theta_{VO}$,
where $\Delta m^2_{i}$ and $\sin^22\theta_{i}$, i=VO,SMA,LMA,
are the values of the two parameters
required by the VO and the SMA and LMA MSW solutions.

    In spite of the strong indications from the 
mean event rate data that the solar neutrinos 
take part in one of the four types of 
oscillations/transitions discussed above, 
none of the physical 
effects considered to be the ``hallmarks'' of the four 
solutions, have been observed so far.
More concretely, i) the specific 
seasonal variation effect 
predicted in the case of the VO solution (see the second 
and the third articles 
quoted in \cite{Pont67} and, e.g., \cite{KP96,Berez98,MaPet99} 
and the references quoted therein) ,
ii) the day-night (D-N) effect 
which should take place
in the case of the MSW solutions (see, e.g., \cite{Hata,BKS98,BalWen,Art2} 
and the references quoted therein)
and iii) the characteristic distortions
of the $e^{-}-$spectrum measured by the 
Super-Kamiokande (SK) collaboration, 
which are predicted 
in the cases of the VO 
and SMA MSW solutions (see, e.g., \cite{KP96,Hata,BKS98}),
have not been observed \cite{Suzuki}.
The unexpected rise of the spectrum at 
$E_{e} \gtap 12.5~{\rm MeV}$
seen in the SK experiment,
$E_e$ being the recoil $e{-}$ total energy,
may indicate an anomalously large 
contribution 
in the above energy range 
from the so-called ``hep'' 
solar neutrinos \cite{BK98}
produced in the reaction
$p + ^{3}He \rightarrow ^{4}He + e^{+} + \nu_e$.
Below 12.5 MeV the measured spectrum is 
compatible with the absence of distortions.
The whole SK data on the $e^{-}-$spectrum, including the points
above 12.5 MeV, seem to favor the VO solution with
\cite{Berez98,Suzuki}
``large'' $\Delta m^2 \cong 4.3\times 10^{-10} {\rm eV}^2$ and
$\sin^22\theta \cong 0.9$, and the LMA MSW solution.
Obviously, more precise data is needed to resolve
these ambiguities and/or to rule out some of the four solutions.

  Very strong evidences for oscillations of the atmospheric 
$\nu_\mu$ and $\bar{\nu}_{\mu}$ neutrinos 
\footnote{We will often use 
the name ``atmospheric $\nu_{\mu}$ (or $\nu_e$)'' 
to indicate both neutrinos and antineutrinos.}   
have been obtained, as is well-known, in the 
SK experiment \cite{Suzuki}. 
These include the measured nonzero Up - Down asymmetry
(a $\sim$ 7 s.d. effect!) 
and the observed substantial 
Zenith angle dependence of
the rates of the sub-GeV and multi-GeV 
$\mu-$like events.
No similar effects were observed 
in the samples of $e-$like events. 
This implies that if the atmospheric $\nu_{\mu}$ 
take part in oscillations, which is the only
plausible explanation of the SK 
$\mu-$like data available at present,
the dominant oscillation modes
should not include the $\nu_e$.
We are therefore left with two possibilities:
the dominant oscillations can be either of the type 
$\nu_{\mu} \leftrightarrow \nu_{\tau}$ 
or $\nu_{\mu} \leftrightarrow \nu_{s}$.
The values of the two-neutrino 
oscillations parameters following (at 90\% C.L.)
from the data are the following \cite{Suzuki}:
$$\nu_{\mu} \leftrightarrow \nu_{\tau}:~~~~
10^{-3}~ {\rm eV}^2\ltap \Delta m^2 \ltap
8.0\times 10^{-2}~ {\rm eV}^2,~~\eqno(5a)
$$
$$~~~~~~~~~~~~~~~~~~~~~
0.86 \ltap \sin^22\theta \leq 1.0,~\eqno(5b)
$$
$$\nu_{\mu} \leftrightarrow \nu_{s}:~~~~
2\times 10^{-3}~ {\rm eV}^2\ltap \Delta m^2 \ltap
7.0\times 10^{-2}~ {\rm eV}^2,~\eqno(6a)
$$
$$~~~~~~~~~~~~~~~~~~~~~
0.86 \ltap \sin^22\theta \leq 1.0.~~~\eqno(6b)
$$
\noindent The $L/E$ or the $L$ (or Zenith angle) independent
suppression of the atmospheric $\nu_{\mu}$ flux, where
$E$ is the neutrino energy and $L$ is the length of the path 
the neutrinos travel before reaching the detector,
are incompatible with the atmospheric neutrino data
\cite{LipLusign}.
In particular, the hypothesis of the 
atmospheric neutrino decay, or of 
gravitationally induced oscillations \cite{GravOsc}
of the atmospheric $\nu_{\mu}$,
are disfavored by the 
SK atmospheric neutrino data
(including the data on the through-going and
stopping muons). 

  Indications for neutrino oscillations 
have been obtained also in the LSND experiment \cite{LSND}: 
the anomalous events observed
by the LSND collaboration can be interpreted as being due to 
$\bar{\nu}_{\mu} \leftrightarrow \bar{\nu}_{e}$ oscillations
with $\Delta m^2$ lying in the interval $\sim$(0.3 - 10)~eV$^{2}$ and
$\sin^22\theta \sim$ few$\times 10^{-3}$. 
The KARMEN collaboration,
which performs a search for the 
indicated type of oscillations
in the same energy range, but at an approximately
two times smaller distance between 
the neutrino source and the detector
than in the LSND experiment,
has not observed anomalous events
in excess of their estimated background \cite{KARMEN}.
The KARMEN results cannot completely rule out, however, 
the possibility that the LSND anomalous events
are due to small mixing 
$\bar{\nu}_{\mu} \leftrightarrow \bar{\nu}_{e}$ oscillations
with values of $\Delta m^2$ in the indicated range,
although they exclude a large fraction of the 
region in the $\Delta m^2 - \sin^22\theta$ plane
suggested by the LSND data.

  An important neutrino oscillation constraint 
has been obtained in the CHOOZ experiment \cite{CHOOZ} 
with reactor $\bar{\nu}_e$. The CHOOZ collaboration 
has not observed a disappearance of the 
$\bar{\nu}_e$ at a distance of $\sim$ 1 km from the reactor.
Interpreted in terms of two-neutrino oscillations
this result implies, e.g., that
$$for~ \Delta m^2 \geq 1.5\times 10^{-3}~ {\rm eV}^2,~~~
\sin^22\theta < 0.22~~(90\%~ C.L.). ~\eqno(7) $$
  
    Rather stringent constraints on the neutrino masses have been 
derived in the $^{3}$H $\beta-$decay experiments as well as 
in the experiments searching for neutrinoless double beta 
($(\beta \beta)_{0\nu}-$) decay, $(A,Z) \rightarrow (A,Z+2) + e^{-} + e^{-}$.
The latter is allowed if the neutrinos with definite 
mass in vacuum are Majorana particles.
The upper limit on the electron neutrino mass, $m(\nu_e)$, 
obtained in the Moscow 
\cite{Lobash} and Mainz \cite{Bonn} 
$^{3}$H $\beta-$decay experiments reads:
$$m(\nu_e) < 2.5~{\rm eV},~~~m(\nu_e) < 2.9~{\rm eV}~~~(95\%~ C.L.).
~~\eqno(8)$$
\noindent There are plans to improve these limits by a factor of
$\sim$ 3, and thus to probe the 1 eV region \cite{Lobash,Bonn}.
The best limit on the effective neutrino mass parameter
$< m _{\nu}>$ extracted form the data 
on the $(\beta \beta)_{0\nu}-$decay (see further),
has been derived in the Heidelberg - Moscow 
$^{76}$Ge experiment \cite{HeidMosc}:
$$  |< m _{\nu}>| ~ < (0.5 - 1.0)~{\rm eV}~~(90\%~ C.L.) .~~~\eqno(9)$$
\noindent The range in eq. (9) reflects the uncertainty in the 
calculations of the corresponding nuclear matrix elements.
This limit is planned to be 
improved at least by a factor of $\sim (3 - 4)$ 
in the ongoing Heidelberg - Moscow experiment and in the
NEMO experiment which is under preparation.
Two experiments have been proposed, 
GENIUS and CUORE \cite{CUORE}, with a projected sensitivity 
to values of $|< m _{\nu}>|$ as small as 
$\sim (5\times 10^{-2} - 10^{-3})~{\rm eV}$.

     Recent developments in the field of 
astrophysics and cosmology suggesting
the existence of a nonzero cosmological constant
imply, in particular,
that the neutrinos with masses
whose sum is $\sim (4 - 6)~{\rm eV}$
are no longer required to provide the
hot dark matter component in 
the Universe \cite{Graciela}.
Nevertheless, neutrinos having a mass
exceeding $\sim$ 1 eV can be 
cosmologically relevant. 
 
   To summarize, the solution of the solar neutrino problem,
the explanation of the atmospheric neutrino data and of the 
LSND results suggest three very different values of the parameter
$\Delta m^2$, namely, $\Delta m^2_{\odot} \ltap 10^{-4}~{\rm eV^2}$,
$\Delta m^2_{atm} \cong (10^{-3} - 10^{-2})~{\rm eV^2}$, 
$\Delta m^2_{LSND} \cong (0.3 - 10.0)~{\rm eV^2}$. 
Correspondingly, we have: 
$\Delta m^2_{\odot} \ll \Delta m^2_{atm} <~(\ll)   
\Delta m^2_{LSND}$. The minimal scheme 
having three independent values of $\Delta m^2$ 
is, obviously, a scheme with four mixed neutrinos.
The LEP data on the number of the light 
neutrinos coupled
to the $Z^{0}-$boson and the cosmological 
constraints on the number of light neutrinos
imply that the needed fourth 
weak-eigenstate neutrino
must be a sterile neutrino, $\nu_s$.   

   In Section 3 we will describe the
solar and atmospheric neutrino oscillations/transitions
in schemes with three-neutrino mixing,
while in Section 
4 we will consider examples of schemes with four-neutrino mixing
which can accommodate also the 
$\bar{\nu}_{\mu} \leftrightarrow \bar{\nu}_{e}$ oscillations 
suggested by the LSND data.
\section{Three-Flavour Neutrino Mixing}
\indent Consider the case of three-flavour neutrino mixing,
$$|\nu_l>~ = \sum_{k=1}^3 U_{lk}^* |\nu_k>, \hspace{1cm} l=e,\mu,\tau, 
~~\eqno(10)$$
\noindent where $|\nu_l>$ is the state vector of
the (left-handed) flavour neutrino $\nu_l$ (with momentum
$\overrightarrow{p}$), $|\nu_k>$ 
is the state vector of a neutrino
$\nu_k$ possessing a definite mass $m_k$ 
(and momentum $\overrightarrow{p}$),
$m_k \neq m_j$, $k \neq j = 1,2,3$, 
$m_1 < m_2 <  m_3$, and U is a $3 \times 3$
unitary matrix -- the lepton mixing matrix. 
It is natural to assume in this case that
one of two independent 
neutrino mass-squared differences,
say, $\Delta m^2_{21}$, 
is relevant for the VO or MSW solutions 
of the solar neutrino problem, 
and has a value 
in one of the intervals given in eqs. (1a) - (4a), 
while $\Delta m^2_{31}$  
is responsible for the dominant 
oscillations of the atmospheric $\nu_{\mu}$,
$\nu_{\mu} \leftrightarrow \nu_{\tau}$.
and lies in the interval (5a). 
For the indicated values of 
$\Delta m^2_{21}$ and $\Delta m^2_{31}$ and  
$$\Delta m^2_{21} = \Delta m^2_{\odot} \ll 
\Delta m^2_{atm} = \Delta m^2_{31}~,~~~\eqno(11)$$
\noindent the relevant three-neutrino VO or MSW transition  
probabilities, describing the solar neutrino 
conversion into another
flavour neutrino, as well as the three-neutrino 
oscillation probabilities for 
the atmospheric neutrinos,
reduce effectively to 
two-neutrino oscillation/transition 
probabilities \cite{ADeR80,CSL87,3nuSP88}. 
We have:
$$P_{\odot}(\nu_e\rightarrow \nu_e) \cong |U_{e3}|^4 + 
(1 - |U_{e3}|^2)^2~P^{2\nu}_{\odot}(\nu_e\rightarrow \nu_e),
~\eqno(12)$$ 
$$P_{atm}(\nu_{\mu}\rightarrow \nu_{\tau}) = 
P_{atm}(\bar{\nu}_{\mu}\rightarrow \bar{\nu}_{\tau}) \cong
2|U_{\mu 3}|^2 |U_{\tau 3}|^2~\left ( 1 -
              \cos {\Delta m^2_{31} \over {2E}}L \right ),~
\eqno(13)$$ 
$$P^{vac}_{atm}(\nu_{\mu} (\nu_e)\rightarrow \nu_{e} (\nu_{\mu})) \cong 
2|U_{\mu 3}|^2 |U_{e 3}|^2~\left ( 1 -
              \cos {\Delta m^2_{31} \over {2E}}L \right ),
~\eqno(14)$$
\noindent $P^{vac}_{atm}(\nu_{\mu} (\nu_e)
\rightarrow \nu_{e} (\nu_{\mu})) \cong
P^{vac}_{atm}(\bar{\nu}_{\mu} (\bar{\nu}_e)
\rightarrow \bar{\nu}_{e} (\bar{\nu}_{\mu}))$,
$$P_{CHOOZ}(\bar{\nu}_{e}\rightarrow \bar{\nu}_{e}) \cong
1 - 2|U_{e3}|^2 (1 - |U_{e3}|^2)~\left (1 -
              \cos {\Delta m^2_{31} \over {2E}}L \right ).~\eqno(15)$$
\noindent Here $P_{\odot}(\nu_e\rightarrow \nu_e)$ is the solar $\nu_e$
survival probability if (10) and (11) hold, 
$P^{2\nu}_{\odot}(\nu_e\rightarrow \nu_e) \equiv 
P^{2\nu}_{\odot}(\Delta m^2_{21}/2E, \sin^22\theta_{12}, |U_{e3}|^2)$ 
is the VO or MSW two-neutrino mixing 
solar $\nu_e$ survival probability, where 
$$\sin^22\theta_{12} = 4 {{|U_{e1}|^2~|U_{e2}|^2}\over {(|U_{e1}|^2 +
|U_{e2}|^2)^2}},~~
\cos 2\theta_{12} =  {{|U_{e1}|^2 - |U_{e2}|^2}\over {|U_{e1}|^2 +
|U_{e2}|^2}},~\eqno(16)$$
\noindent and the other notations are self-explanatory.
In the case of the VO solution  
the probability $P^{2\nu}_{\odot}(\nu_e\rightarrow \nu_e)$ 
does not depend on $|U_{e3}|^2$ and is given by the standard 
two-neutrino mixing expression with 
$\Delta m^2_{21}$ and $\sin^22\theta_{12}$ playing the 
role of the two oscillation parameters.
If the solar $\nu_e$ take part in MSW transitions, 
the dependence of $P^{2\nu}_{\odot}(\nu_e\rightarrow \nu_e)$
on $|U_{e3}|^2$ amounts to the change of the matter term
\cite{CSL87,3nuSP88}
$$\sqrt{2}G_{F}N_e \rightarrow 
\sqrt{2}G_{F}N_e(1 - |U_{e3}|^2),~~\eqno(17)$$  
\noindent $N_e$ being the electron number density,
in the standard expression for the two-neutrino mixing
survival probability. 
Let us note finally that the expression for the 
probability
$P^{vac}_{atm}(\nu_{e}\rightarrow \nu_{\tau}) = 
P^{vac}_{atm}(\bar{\nu}_{e}\rightarrow \bar{\nu}_{\tau})$ can 
be obtained from the expression in the right-hand 
side of eq. (13) by replacing 
$|U_{\mu3}|^2$ with $|U_{\tau 3}|^2$.

  Under the condition (11), the solar 
neutrino survival probability (12) depends only on 
the elements of the first 
row of the lepton mixing matrix, i.e., on $|U_{ei}|^2$, i=1,2,3, while
the oscillations of the atmospheric $\nu_{\mu}$ and 
$\nu_e$ are controlled by the elements of the third column
of $U$, $|U_{l3}|^2$, $l=e,\mu,\tau$. 
The other elements of $U$ are not 
accessible to direct experimental determination.

     The CHOOZ limit (7) and analysis of the
solar neutrino data based on  eq. (12)
imply that $|U_{e3}|^2$ has to be small: for 
$\Delta m^2_{31} \gtap 1.5\times 10^{-3}~{\rm eV^2}$ one has
$$|U_{e3}|^2 \ltap 0.05.~~\eqno(18)$$
\noindent This may be indicating that $|U_{e3}| \ll 1$, or even 
that $|U_{e3}| \cong 0$. The lepton mixing matrix takes 
the particularly simple form of {\it bi-maximal mixing}
(see, e.g., \cite{BGG98,Altarelli} and 
the references quoted therein) 
in the case of the VO solution of the solar neutrino 
problem if $|U_{e3}| \cong 0$ and we assume that
$\sin^22\theta_{12} = 1$ and $|U_{\mu 3}|^2 = |U_{\tau 3}|^2$:
$$U \cong
\left(\begin{array}{ccc} 
{1\over {\sqrt{2}}} & {1\over {\sqrt{2}}} & 0\\
- {1\over 2} & {1\over 2} & {1\over {\sqrt{2}}}\\
{1\over 2} & -{1\over 2} & {1\over {\sqrt{2}}}\\ 
\end{array} \right).~~\eqno(19)$$
\noindent In this case there will be no CP-violation 
in the oscillations of neutrinos in vacuum;
and if the neutrinos with definite mass $\nu_i$, $i=1,2,3,$
are Dirac particles there will be no CP-violation at all 
in the lepton sector.
If, however, $\nu_i$ are massive Majorana neutrinos,
there can be CP-violation related effects 
in processes which
are associated with the Majorana nature of the 
$\nu_i$'s, as like the  
$(\beta \beta)_{0\nu}-$decay \cite{BPet87,BGKP96}, etc.
The solar $\nu_e$ will oscillate
with equal probabilities into $\nu_{\mu}$ and $\nu_{\tau}$ and the 
atmospheric $\nu_e$ will not oscillate over the distances 
probed by the atmospheric neutrino experiments 
($L \ltap 12800~$km).

   The above conclusions will be approximately valid
if $|U_{e3}| \neq 0$, but $|U_{e3}| \ll 1$.
In particular, the CP-violation effects 
in neutrino oscillations
will be strongly suppressed. Further, one can 
determine the elements of $U$ 
from the analysis 
of the solar and atmospheric neutrino data.
There are three possible solutions of the solar neutrino problem
- the MSW $\nu_e \rightarrow \nu_s$
transition solution cannot be realized 
in the case of (10).
Utilizing a standard parametrization of $U$ 
we have:   
$$U =
\left(\begin{array}{ccc} 
c_{12}c_{13} & s_{12}c_{13} & U_{e3}\\
- s_{12}c_{23} - c_{12}s_{23}U_{e3}^{*} & 
 c_{12}c_{23} - s_{12}s_{23}U_{e3}^{*} & s_{23}c_{13}\\
s_{12}s_{23} - c_{12}c_{23}U_{e3}^{*} & -c_{12}s_{23} - s_{12}c_{23}U_{e3}^{*}
& c_{23}c_{13}\\ 
\end{array} \right)$$
$$\cong 
\left(\begin{array}{ccc} 
c_{12}c_{13} & s_{12}c_{13} & << 1\\
- s_{12}c_{23} & c_{12}c_{23} & s_{23}\\
s_{12}s_{23} & -c_{12}s_{23} & c_{23}\\ 
\end{array} \right),\eqno(20)$$
\noindent where $c_{ij} \equiv \cos \theta_{ij}$,
$s_{ij} \equiv \sin \theta_{ij}$ and 
$U_{e3} = s_{13}e^{-i\delta_{13}}$, $\delta_{13}$ being the 
Dirac CP-violation phase and we have not 
written explicitly the possible
Majorana CP-violation phases (see, e.g., \cite{BPet87}). 
The angles $\theta_{12}$ and $\theta_{23}$ in (20) 
are fixed (with a known ambiguity)
within rather narrow ranges by the
solar and the atmospheric neutrino 
data and this determines all the
elements of the lepton mixing matrix in eq. (20). One finds for 
the three solutions of the solar neutrino problem:
$$~~~~~~~~~~~~~~VO:~~~~~|s_{12}| \cong 0.48 - 0.71,~~~\eqno(21a)$$
$$SMA~ MSW:~~~~~s_{12} \cong 0.02 - 0.05,~~~\eqno(21b)$$
$$LMA~ MSW:~~~~~s_{12} \cong 0.30 - 0.55.~~~\eqno(21c)$$
\noindent The atmospheric neutrino data implies:
$$\nu_{\mu}\leftrightarrow \nu_{\tau}:~~|s_{23}| 
\cong 0.50 - 0.71.~~~\eqno(22)$$   
\noindent Clearly, the lepton mixing matrices  
corresponding to eqs. (21a) - (22) are  
very different from the quark mixing matrix.

   It follows from the above discussion that 
under the condition (11),
the magnitude of $|U_{e3}|$ 
controls, in particular, the 
$\nu_e \leftrightarrow \nu_{\mu,\tau}$ and 
$\nu_{\mu} \leftrightarrow \nu_{e}$ oscillations
of the atmospheric and terrestrial (i.e., ``man made'')
neutrinos as well as the magnitude of the 
CP-violation effects in
the oscillations of neutrinos 
and in the lepton sector, in general. It would be 
extremely important 
to obtain better experimental limits on, 
or determine the value of
$|U_{e3}|$. The long base-line 
neutrino oscillations experiments MINOS and ICARUS  
\cite{Litchf} are envisaged to be sensitive to values of
$|U_{e3}|^2$ as small as $\sim 5\times 10^{-3}$.

  There is an additional very interesting
new physical effect related to the $|U_{e3}|$.
We have notice that the latter ``drives'' 
the sub-dominant 
$\nu_e \leftrightarrow \nu_{\mu,\tau}$ and 
$\nu_{\mu} \leftrightarrow \nu_{e}$ oscillations
of the atmospheric $\nu_{\mu}$ and
$\nu_e$. As was pointed out in \cite{SP1}, 
for the neutrinos passing through the Earth,
these oscillations can be strongly amplified  
by a new resonance-like mechanism, which 
differs from the MSW one 
and takes place when the 
neutrinos cross the Earth core.
The new mechanism can cause a total 
neutrino conversion \cite{chpet99}.
At small mixing angles ($\sin^22\theta \ltap 0.10$),
the maxima due to this
new enhancing mechanism in 
$P(\nu_{\mu} \rightarrow \nu_{e})$ and
$P(\nu_e \rightarrow \nu_{\mu(\tau)})$ 
are absolute maxima and dominate in 
these probabilities:
they are considerably 
larger than the local maxima
of $P(\nu_{\mu} \rightarrow \nu_{e})$ and
$P(\nu_e \rightarrow \nu_{\mu(\tau)})$,
associated with the MSW effect
taking place in the Earth core (mantle) 
\footnote{The effect of the new enhancement
is less dramatic at large mixing angles.
The enhancement is present and dominates
also in the $\nu_2 \rightarrow \nu_{e}$ transitions
in the case of $\nu_e - \nu_{\mu (\tau)}$ or
$\nu_e - \nu_{s}$ mixing and
in the $\nu_e \rightarrow \nu_{s}$ and
$\bar{\nu}_{\mu} \rightarrow \bar{\nu}_{s}$
transitions at small mixing angles
\cite{SP1,chpet99,s5398,SPNewE98}.}.
The mixing angle 
which is relevant for 
the magnitude of the enhancement, 
is determined 
in the case of interest by 
$|U_{e3}|^2$ (see further). 
Even at small mixing angles
the enhancement
is relatively
wide \cite{SP1,chpet99,s5398,SPNewE98}
in the Nadir (or Zenith) angle \footnote{The Nadir angle
determines uniquely the neutrino trajectory through
the Earth.}, $h$, and in the neutrino energy $E$ -
it is somewhat wider than the MSW resonance, 
and therefore can produce observable effects.
As was shown in \cite{chpet99}, 
the new mechanism of 
enhancement of the neutrino transitions in the Earth
is of interference nature: it is 
caused by a maximal 
constructive interference
between the probability amplitudes
of the neutrino transitions in the 
Earth mantle and in the Earth core.
Thus, the effect has nothing to do 
with the parametric resonance in the 
neutrino transitions,
discussed in \cite{Param} and  
possible in a medium with
periodic change of density.

 The conditions for a total neutrino 
conversion due to the new effect,
$P(\nu_{\mu} \rightarrow \nu_e) = 1$,
in the two-neutrino mixing case
include specific relations between
the phase differences the 
neutrino energy-eigenstates
acquire after crossing the Earth mantle,
$2\phi'$, and the core, $2\phi''$
and the mixing angles in matter in the
mantle, $\theta_m'$, and in the core, 
$\theta_m''$.
For the two-neutrino 
$\nu_{\mu (e)} \rightarrow \nu_{e (\mu ;\tau)}$,
$\nu_e \rightarrow \nu_{s}$ and
$\bar{\nu}_{\mu} \rightarrow \bar{\nu}_{s}$
transitions ($\Delta m^2~\cos 2\theta > 0$) they read \cite{chpet99}:
$$
\tan\phi' = \pm\sqrt{{\displaystyle -\cos 2\theta_m''\over
\displaystyle\cos(2\theta_m''- 4\theta_m')}},~~
\tan\phi'' = \pm{\displaystyle \cos 2\theta_m'\over \sqrt{
\displaystyle-\cos(2\theta_m'')\cos(2\theta_m''- 4\theta_m')}},
~\eqno(23)$$
\noindent where the signs are correlated
\footnote{The conditions for the
$\nu_2 \rightarrow \nu_{e}$ transitions
can formally be obtained from eq. (23)
by replacing $2\theta_m''$ and $2\theta_m'$ in the
expressions for $\phi'$ and $\phi''$
with ($2\theta_m'' - \theta$) and ($2\theta_m' - \theta$).}.
It is quite remarkable that 
these conditions are satisfied \cite{chpet99,SP1}
for the Earth-core-crossing 
solar and atmospheric neutrinos.

  The $\nu_{e} \rightarrow \nu_{\mu ;\tau}$ and 
$\nu_{\mu} \rightarrow \nu_{e}$ transition probabilities
in the Earth are given under the conditions (10) - (11) by 
(see, e.g., \cite{3nuSP88}):
$$P^{3\nu}_{E}(\nu_{\mu} (\nu_e)
\rightarrow \nu_{e} (\nu_{\tau}))
\cong {|U_{\mu (\tau) 3}|^2\over{1 - |U_{e3}|^2}}~
P^{2\nu}_{E}(\Delta m^2_{31}, \sin^22\theta_{13}),
$$
\noindent where $P^{3\nu}_{E}(\nu_{e} 
\rightarrow \nu_{\mu}) \cong P^{3\nu}_{E}(\nu_{\mu} 
\rightarrow \nu_{e})$,
$P^{2\nu}_{E}(\Delta m^2_{31}, \sin^22\theta_{13})$
is a known two-neutrino transition probability for the
Earth-core-crossing neutrinos \footnote{An analytic expression
for $P^{2\nu}_{E}(\Delta m^2_{31}, \sin^22\theta_{13})$ can be found
in \cite{SP1}.} and 
$\sin^22\theta_{13} \equiv 4|U_{e3}|^2(1 - |U_{e3}|^2)$.
For the fluxes of the atmospheric 
$\nu_{e,\mu}$ with energy $E$,
crossing the Earth along a trajectory with Zenith angle
$\theta_{z}$ before reaching the detector
we get 
\cite{s5398,SPNewE98}:
$$ \Phi_{\nu_e} \cong \Phi^{0}_{\nu_e} \left ( 1 +
  [s^2_{23}~r(E,\theta_{z}) - 1]~
P^{2 \nu}_{E}(\Delta m^2_{31}, \sin^22\theta_{13}) \right ),~\eqno(24)$$
$$\Phi_{\nu_{\mu}} \cong \Phi^{0}_{\nu_{\mu}}  ( 1 +
 s^4_{23}~ [(s^2_{23}~r(E,\theta_{z}))^{-1} - 1]~
P^{2\nu}_{E}(\Delta m^2_{31}, \sin^22\theta_{13})~~~~~~~~~~~$$
$$~~~~~~~~~~~~~~~~~~~~~~~~~-  2c^2_{23}s^2_{23}~[ 1 -
Re~( e^{-i\kappa}
A^{2\nu}_{E}(\Delta m^2_{31}, \sin^22\theta_{13}))] ),~ \eqno(25)$$
\noindent where 
$\Phi^{0}_{\nu_{e(\mu)}} = \Phi^{0}_{\nu_{e(\mu)}}(E,\theta_{z})$ is the
$\nu_{e(\mu)}$ flux in the absence of 
oscillations, $s^2_{23} \equiv |U_{\mu 3}|^2/(1 - |U_{e3}|^2)$, 
$r(E,\theta_{z}) \equiv \Phi^{0}_{\nu_{\mu}}/
\Phi^{0}_{\nu_{e}}$,
$A^{2\nu}_{E}(\Delta m^2_{31}, \sin^22\theta_{13})$
is a known amplitude 
of two-neutrino transitions in the Earth, and 
$\kappa \equiv \kappa(\Delta m^2_{31}, \sin^22\theta_{13})$ 
is a known phase factor. Analytic expressions 
for $A^{2\nu}_{E}$ and  
$\kappa$ are given in \cite{s5398,SPNewE98}.

   The probability 
$P^{2\nu}_{E}(\Delta m^2_{31}, \sin^22\theta_{13})$
can be strongly enhanced to values $\sim 1$ by 
the new resonance-like mechanism.
At $\sin^22\theta_{13}\ltap 0.2$ 
and for values of  
$\Delta m^2_{31}$
suggested by the SK data,
the enhancement takes place \cite{SP1} for the  
$\nu_e$ and $\nu_{\mu}$ with 
energies $\sim (1.0 - 10.0)~{\rm GeV}$, which
contribute either to the sub-GeV or to the
multi-GeV $e-$like and $\mu-$like SK
event samples. At small mixing angles
the enhancement holds
practically for all neutrino
trajectories through the Earth core.
The new effect can produce
an excess of e-like events
in the region $-1 \leq \cos\theta_{z}\ltap -0.8$,
$\theta_{z}$ being the Zenith angle,
in the multi-GeV  (or a smaller one - in the
sub-GeV) sample of events.
Actually, the effect may have already
manifested itself  \cite{SP1,s5398,SPNewE98}, 
producing at least part of the
strong Zenith angle dependence observed in the
multi-GeV $\mu-$like event rate in the SK experiment
\footnote{The new effect should also be present in the
$\bar{\nu}_{\mu} \leftrightarrow \bar{\nu}_{s}$
transitions of the atmospheric
multi-GeV  $\bar{\nu}_{\mu}$'s both at small,
intermediate and large mixing angles, 
if the atmospheric neutrinos 
undergo such transitions.}.
The multi-GeV SK data can be used to obtain 
information on the value of 
$\sin^22\theta_{13}$ and thus
of $|U_{e3}|^2$.

 Let add finally that the new enhancement
mechanism is operative also in the transitions
of the Earth-core-crossing solar neutrinos \cite{SP1,chpet99}.
This has important implications
for the tests of the MSW
$\nu_e \rightarrow \nu_{\mu (\tau)}$
transition solutions of the solar neutrino problem, 
as discussed in detail in
\cite{SP1,Art2}.

\section{Scenarios with Four-Neutrino Mixing}

In the case of four-neutrino mixing we have 
$$|\nu_\alpha>~ = \sum_{k=1}^4 U_{lk}^* |\nu_k>, 
\hspace{1cm} \alpha=e,\mu,\tau,s 
~~\eqno(26)$$
\noindent where $U$ is now a $4\times 4$ unitary matrix.
This may seem to be a rather messy case with a 
large number of possible relations between the
three independent neutrino mass-squared differences 
and a large number of mixing parameters: 
the matrix 
$U$ contains now 
6 mixing angles and 3 Dirac type CP-violation phases. 
However, as was shown in \cite{Giunti}, only two possibilities,
in what regards the relations between the different 
$\Delta m^2$, are compatible with the existing data,
including the LSND result and the 
constraints on the neutrino
oscillations parameters obtained in 
the accelerator experiments:
$$\Delta m^2_{43} = \Delta m^2_{\odot} \ll 
 \Delta m^2_{21} = \Delta m^2_{atm}~,~~~\eqno(A)$$ 
\noindent with $\Delta m^2_{LSND} = 
\Delta m^2_{41} \cong \Delta m^2_{42} \cong
\Delta m^2_{31}\cong \Delta m^2_{32}$, and \cite{Barger98}
$$\Delta m^2_{21} = \Delta m^2_{\odot} \ll 
 \Delta m^2_{43} = \Delta m^2_{atm}~,~~~\eqno(B)$$ 
\noindent with $\Delta m^2_{LSND} = 
\Delta m^2_{41}$, etc.
The 
matrix $U$ takes a particularly simple form
if one implements the 
{\it standard} Big Bang Nucleosynthesis (BBN)
constraint on the number of light neutrinos, $N_{\nu}$:
$N_{\nu} \ltap 4$. One finds \cite{Giunti} in the case (A):
$$U \cong 
\left(\begin{array}{cccc} 
0 & 0  & \cos \beta & \sin\beta\\
\cos\gamma & \sin\gamma & 0 & 0 \\
-\sin \gamma & \cos \gamma & 0 & 0\\ 
0 & 0  & - \sin \beta & \cos \beta\\
\end{array} \right)
~~~\eqno(27)
$$
\noindent where $\beta$ and $\gamma$ are determined
by the solar and atmospheric neutrino data, respectively.
The small mixing responsible for the
LSND effect can be accounted for by a 
minor modification
of the matrix (27). The solar neutrino problem is solved
in the scheme (A) by SMA MSW $\nu_e \rightarrow \nu_s$
transitions, while the dominant oscillations of the
atmospheric $\nu_{\mu}$ are of the type
$\nu_{\mu}\leftrightarrow \nu_{\tau}$: the 
$\nu_{\mu}\leftrightarrow \nu_{s}$ 
transitions are strongly suppressed.  
The mixing matrix in the case (B) has a similar structure:
it can be obtained from eq. (27) by interchanging
the first (second) and the third (fourth) columns.
The solutions of the solar and atmospheric neutrino problems
are the same.

\section{The Neutrino Mass Spectrum}
 
\indent The neutrino oscillation experiments 
or the interpretation of
a given set of data (solar, atmospheric, LSND) 
in terms of neutrino
oscillations or MSW transitions, 
does not provide information 
about the absolute values of the neutrino masses.
In the scheme with three-neutrino mixing, for example,
there are three possible types of neutrino mass spectrum
and all of them lead to the same neutrino oscillation phenomenology.
We can have a hierarchical spectrum,
$$m_1 \ll m_2 \ll m_3,~~\eqno(28)$$
\noindent there could be two quasi-degenerate neutrinos,
$$m_1 \cong m_2 \ll m_3,~~or~~m_1 \ll m_2 \cong m_3,~~\eqno(29)$$
\noindent or three quasi-degenerate neutrinos,
$$m_1 \cong m_2 \cong m_3.~~\eqno(30)$$
\noindent In the cases (28) - (29) 
one has to identify 
$\Delta m_{31}$ with $\Delta m^2_{atm}$ in order to 
explain the atmospheric neutrino
data. Correspondingly, we 
have $m_3 \cong \sqrt{\Delta m^2_{atm}}$
and all neutrino masses 
cannot exceed the value $\sqrt{\Delta m^2_{atm}}$: 
$m_i \ltap$~(0.03 - 0.10) eV. The values of 
the neutrino masses are too small to be observed 
in the direct search experiments, 
like the $^{3}$H $\beta-$decay
experiments.
    
     The situation is very different if the 
three massive neutrinos are 
quasi-degenerate in mass, eq. (30). 
The massive neutrinos can be 
cosmologically relevant: one can have
$m_i \cong {\rm few~eV} \gg \sqrt{\Delta m^2_{atm}}$, i=1,2,3.
Actually, the experimental upper limit (9) implies in this case
 $m_i < (2.5 - 3.0)~{\rm eV}$. 
If the future $^{3}$H $\beta-$decay
experiments will observe an effect of nonzero
neutrinos mass $\sim (1 - 3)$ eV, that would imply 
within the scheme (10)
that the three massive neutrinos are 
quasi-degenerate. In the four-neutrino mixing 
schemes discussed in Section (4),
neutrino masses in the range $\sim (1 - 3)$ eV 
are possible \cite{BGG98,Giunti} in 
the scheme (A), but not in the scheme (B).
A negative result of 
the $^{3}$H experiments will not provide
an information on the structure 
of the neutrino mass spectrum.    

  Additional information on the neutrino 
mass spectrum can be obtained 
in the future high sensitivity  
$(\beta \beta)_{0\nu}-$decay experiments, provided
the massive neutrinos are Majorana particles
\cite{KPR84}.
For $m_i \ll$ MeV, i=1,2.3,4,
which is the case of interest,
we have for the neutrino mass 
parameter measured in
the $(\beta \beta)_{0\nu}-$decay experiments 
(see, e.g., \cite{BPet87}):
$$< m _{\nu}> = \sum_{k=1}
m_k \eta_k~|U_{ek}|^{2},~~\eqno(31)$$
\noindent where $i\eta_k =\pm i$ is the CP-parity
of the Majorana neutrino $\nu_k$ and 
we have written for simplicity  
the expression for 
$< m _{\nu}>$ in the case of 
CP-conservation. Because the 
massive Majorana neutrinos
can have opposite CP-parities, 
there can be cancellation
between the different terms in the sum in eq. (31).

   We can use the values of the lepton mixing angles
and of the masses $m_k$ in the cases (28) - (30) required
by the three possible solutions of the solar
neutrino problem and the neutrino oscillation explanation of the
atmospheric neutrino data together with the CHOOZ limit (7) 
to derive upper bounds on the value of 
$|< m _{\nu}>|$. One arrives in this way to 
the following conclusions \cite{KPR84}. 
If the spectrum is hierarchical (28),
then $|< m _{\nu}>| \leq 0.008~$eV.
In the case of two quasi-degenerate neutrinos (28) or (29), 
$|< m _{\nu}>|$ can be larger than 0.01 eV, but cannot exceed
0.10 eV. Finally, if (30) is realized, 
$|< m _{\nu}>|$ can have a value in the interval 
(0.10 - 1.0) eV. The latter is valid also in the case of 
4-neutrino mixing in the scheme (A), while in scheme (B)
one has $|< m _{\nu}>| < 0.01~$eV \cite{BGG98,Giunti}.
Thus, if the $(\beta \beta)_{0\nu}-$decay will be observed in the
future experiments, the measurement of its rate can provide 
important information on the neutrino
mass spectrum.

\section{Conclusions} 
To summarize, the possible patterns of 
the lepton mixng are emerging
from the analyses of the solar 
and atmospheric neutrino data.
They are very different from the 
pattern of the quark mixing.
Future $^{3}$H $\beta-$decay 
and $(\beta \beta)_{0\nu}-$decay
experiments can provide information on 
the neutrino mass spectrum.
However, more data is needed to 
establish i) the true cause of the solar neutrino
deficit (VO, or MSW transitions, or may be something else...),
and ii) the type of dominant oscillations of the atmospheric 
$\nu_{\mu}$ ($\nu_{\mu} \rightarrow \nu_{\tau}$ or
$\nu_{\mu} \rightarrow \nu_{s}$).
It is also very important to establish whether 
the sub-dominant $\nu_{\mu} \rightarrow \nu_{e}$
and $\nu_{e} \rightarrow \nu_{\mu,\tau}$ oscillations of the
atmospheric neutrinos take place. These sub-dominant 
oscillations
and the transitions of the 
solar neutrinos in Earth can be
strongly (maximally) enhanced by a new 
type of mechanism
which differs from the MSW one 
and takes place when the neutrinos cross
the Earth core on the way to the detector.
It would be quite remarkable to observe experimentally
the indicated enhancement.

    We believe at least some of the 
above questions will be answered
by the future experiments: 
SNO, BOREXINO, ICARUS, HERON,  
etc. in the case of solar neutrinos, 
and by K2K, MINOS, KAMLAND, mini-BOONE, etc. 
  The future in the field of 
physics of neutrino 
oscillations and  massive neutrinos 
looks as exciting as the present.    

\end{document}